\documentclass{emulateapj}
\usepackage{apjfonts}
\newcommand{\scaleup}{\epsscale{1.1}}

\newcommand{\plotterr}{\plotone}
\newcommand{\breaker}{}

\newcommand{\etal}{et al.}

\newcommand{\msun}{M_{\sun}}

\newcommand{\fgas}{f_{\rm gas}}

\shorttitle{Satellite Orbits and Disk Heating}
\shortauthors{Hopkins \etal}
\slugcomment{Submitted to ApJ, June 13, 2008}
\begin{document}

\title{The Radical Consequences of Realistic Satellite Orbits for 
the Heating and Implied Merger Histories of Galactic Disks} 
\author{Philip F. Hopkins\altaffilmark{1}, 
Lars Hernquist\altaffilmark{1}, 
Thomas J. Cox\altaffilmark{1,2}, 
Joshua D. Younger\altaffilmark{1}, 
\&\ Gurtina Besla\altaffilmark{1}
}
\altaffiltext{1}{Harvard-Smithsonian Center for Astrophysics, 
60 Garden Street, Cambridge, MA 02138}
\altaffiltext{2}{W.~M.\ Keck Postdoctoral Fellow at the 
Harvard-Smithsonian Center for Astrophysics}

\begin{abstract}

Previous models of galactic disk heating in interactions invoke restrictive
assumptions not necessarily valid in modern $\Lambda$CDM contexts: that satellites are
rigid and
orbits are circular, with slow decay over many orbital periods from
dynamical friction. This leads to a linear scaling of disk heating with
satellite mass: disk heights and velocity 
dispersions scale $\propto M_{\rm sat}/M_{\rm disk}$. In
turn, observed disk thicknesses present strong constraints on merger histories:
the implication for the Milky Way is that $<5\%$ of its mass could come from
mergers since $z\sim2$, in conflict with cosmological predictions. More
realistically, satellites merge on nearly radial orbits, and once near the disk,
resonant interactions efficiently remove angular momentum while tidal
stripping removes mass, leading to rapid merger/destruction in a couple of free-fall
plunges. Under these conditions the proper heating efficiency is non-linear in
mass ratio, $\propto (M_{\rm sat}/M_{\rm disk})^{2}$. 
We derive the scaling of disk scale heights and
velocity dispersions as a function of mass ratio and disk gas content in this
regime, and show that this accurately describes the results of simulations with
appropriate ``live'' halos and disks. Under realistic circumstances, we show 
that disk heating in
minor mergers is suppressed by an order of magnitude relative to the expectations of
previous analyses. We show that the Milky Way disk could have 
experienced $\sim5-10$ 
independent 1:10
mass-ratio mergers since $z\sim2$, in agreement with cosmological models. 
Because the realistic heating rates are non-linear in mass, the predicted heating is 
dominated by the more stochastic, rare low mass-ratio mergers, and the existence 
of populations with little or no thick disk does not require fundamental 
modifications to the cosmology. This also leads to important differences in 
the predicted isophotal shapes of bulge-disk systems along the Hubble sequence. 

\end{abstract}

\keywords{galaxies: formation --- galaxies: evolution --- galaxies: active --- 
galaxies: spiral --- cosmology: theory}

\section{Introduction}
\label{sec:intro}

In the widely-accepted $\Lambda$CDM cosmology, it is expected that
mergers and interactions between galaxies are crucial to galaxy
assembly because structure grows hierarchically
\citep[e.g.][]{whiterees78}.  On this basis, it has been proposed that
spheroids are produced when galaxies merge \citep[making bulges in
disks and elliptical galaxies;][]{toomre72,toomre77}.  Both theory
\citep[][and references
therein]{ostrikertremaine75,maller:sph.merger.rates,stewart:mw.minor.accretion,
fakhouri:halo.merger.rates} and observation
\citep[e.g.][]{lin:merger.fraction,woods:tidal.triggering,barton:triggered.sf,
woods:minor.mergers} indicate that ``minor''
mergers of mass ratios $\lesssim$\,1:10 are commonplace, and that a
large fraction ($\sim1/2$) of the $\sim L_{\ast}$ galaxy population
has experienced a ``major'' merger (mass ratio $\lesssim$\,1:3) since
$z\sim2-3$ \citep{bell:merger.fraction,
bridge:merger.fractions,kartaltepe:pair.fractions,
lotz:merger.fraction,lin:mergers.by.type}.

The ubiquity of mergers has raised the possibility that cold, thin
disks of galaxies may not survive to $z=0$ in hierarchical models.
From numerical simulations \citep[e.g.][]{barnes:disk.halo.mergers,barnes:disk.disk.mergers,
hernquist:bulgeless.mergers,hernquist:bulge.mergers} it has been known for some time that major
mergers destroy {\it stellar} disks entirely and cause a large
fraction of disk gas to lose angular momentum through gravitational
torques \citep[e.g.][]{barnes.hernquist.91,barneshernquist96} together
yielding dense, quasi-spherical bulges \citep[at least in typical
situations such at those at low redshift, where the gas fractions are
not large; see e.g.][]{hopkins:disk.survival}.  Moreover, although
minor mergers (at least those with mass ratios $\lesssim$\,10:1) likely
do not destroy disks, they are almost an order of magnitude more
frequent than major mergers and as such may pose an even more severe
problem for disk survival.

In the $\Lambda$CDM cosmology it is unlikely than any disk with a
significant stellar age has existed $\sim5-10$\,Gyr without
experiencing a merger of mass ratio 10:1 or larger.  Analytic
arguments \citep{ostrikertremaine75,tothostriker:disk.heating,
sellwood:resonant.disk.thickening} and early generations of numerical
simulations
\citep{quinn.84,quinn86:dynfric.on.sats,
hernquist.89,quinn93.minor.mergers,
hernquist.mihos:minor.mergers,
walker:disk.fragility.minor.merger,naab:minor.mergers} indicate that
gas-poor minor mergers can convert a large fraction of a stellar disk
into bulge and dynamically heat the disk. 
The observed thinness of galactic disks suggests that 
this may be a severe problem: 
\citet{tothostriker:disk.heating} argued that large disks such as 
that in the Milky Way could not have undergone a merger of 
mass ratio $\lesssim10:1$ in the last $\sim10\,$Gyr. More 
recently e.g.\ \citet{gilmore02:last.mw.merger.from.thick.disk,wyse06:thick.disk.stars} 
and \citet{stewart:mw.minor.accretion} emphasized that the tension
between these constraints and the expectation in CDM models that a
number of such mergers should occur implies either a problem with our
understanding of hierarchical disk formation or a challenge to the
concordance cosmological model.

Given the many successes of the $\Lambda$CDM model and observations
that disks do experience (and survive) a large number of mergers, it
seems likely that the main uncertainty is our understanding of disk
galaxy evolution.  This has focused attention on the problem of
forming disks in a cosmological context.  A number of studies have
argued that detailed modeling of gas physics, star formation, and
feedback, along with high
numerical resolution, are required to simulate disk formation and understand 
how disks survive their violent merger histories \citep{weil98:cooling.suppression.key.to.disks,
sommerlarsen99:disk.sne.fb,sommerlarsen03:disk.sne.fb,
thackercouchman00,thackercouchman01,abadi03:disk.structure,
governato04:resolution.fx,governato:disk.formation,
robertson:cosmological.disk.formation,
okamoto:feedback.vs.disk.morphology,scannapieco:fb.disk.sims}. 
But there has been relatively less focus on whether these simulations
produce disks -- even where the disks avoid being converted into
bulges -- that are sufficiently thin relative to observational
constraints. Interestingly, more recent attempts to model disk heating 
in a cosmological context \citep[see e.g.][]{benson:heating.model,
kazantzidis:mw.merger.hist.sim} suggest that disk heating may be 
significantly less efficient than has been assumed. 

Indeed, a closer examination of this question reveals that much of the
conventional wisdom and the implied constraints on the merger history
of massive disks (and the Milky Way in particular) is fundamentally
premised on historical assumptions that do not apply given what we now
know about the $\Lambda$CDM framework.  For example, implicit in the
analysis of \citet{tothostriker:disk.heating} and others is the
assumption that the orbits of merging satellites can be described as
being nearly circular, slowly spiraling in owing to dynamical
friction, while the satellite remains relatively intact. Early
simulations often adopted these choices -- modeling the
satellites as rigid point masses on almost perfectly circular orbits,
including relatively low-mass and smooth dark matter halos (such that
the background density is low, and the in-spiral of the satellite
therefore relatively slow and without much tidal stripping), and/or
using rigid (as opposed to full N-body or ``live'') background
disk/bulge/halo potentials. Moreover, these calculations were almost
always dissipationless, and did not allow for the role of gas in
making the system more robust against mergers and heating.  Under
these conditions, it is straightforward to show that the disk heating
owing to a merger is linear in the mass of the satellite.  This
fundamentally underlies the constraint above: because the scaling is
linear, it is possible to transform an observed disk thickness or
velocity dispersion into an integral constraint on the allowed merger
history of a galaxy.

Given cosmological constraints and numerical capabilities at the time,
these were reasonable assumptions.  However, with improvements in
simulations, yielding the ability to perform true cosmological
simulations with resolved subhalo populations, and 
better understanding of cosmological
parameters and estimates of halo masses and sizes, it is clear that
the assumptions underlying these early works are not appropriate.  For
example, it is now well-established that realistic cosmological orbits
are quite radial, not circular. This is true on all mass
scales, from large dark matter substructures which populate halos
\citep{vandenbosch:no.orbital.circularization.due.to.dyn.frict,
benson:cosmo.orbits,khochfar:cosmo.orbits} to unresolved background
velocity flows in the diffuse stellar and dark matter halo
\citep{cole.lacey:halo.structure,abadi06:stellar.halos}.  Given that
realistic galaxy halos exhibit considerable substructure with large
subhalo populations, it is generally true that systems which pass near
a disk -- those of interest for mergers and disk heating -- will be
those which can transfer angular momentum and collide with the disk on
highly radial passages, rather than circular in-spirals
\citep[][]{font:sats.on.large.orbits,benson:heating.model}.

Moreover, properly accounting for live halos and disks leads to
qualitatively different angular momentum transfer in the final stages
of galaxy mergers: once a satellite reaches a radius comparable to the
disk, resonances between the orbital motion and internal motions of
the disk and halo remove angular momentum efficiently and lead to a
rapid, radial coalescence in a couple of dynamical times
(rather than the large number of orbital times required for a
dynamical friction in-spiral in the case of mass ratios
$\gtrsim$\,1:10).  Furthermore, allowing halos to have the large masses
expected in modern cosmologies and resolving disks with N-body approaches,
with proper live secondaries, leads to efficient tidal stripping of the
satellite once it begins to pass through a disk, transferring orbital
energy to the central bulge and extended dark matter halo
(preferentially the hottest components of the system, rather than the
disk).

The historical picture of ``sinking'' satellites on slowly decaying,
circular orbits is based in part on an old view of the orbits of the
Large and Small Magellanic Clouds (LMC and SMC). 
Recent observations, however, indicate that the LMC and SMC are on a
high-velocity passage near pericenter and are only marginally bound to
the Milky Way \citep{kallivayalil:smc.proper.motion,
kallivayalil:lmc.proper.motion, piatek:kallivayalil.reanalysis}.
\citet{besla:mc.orbits} used these results and modern cosmological
models of the Milky Way halo to infer their orbital motion, and argued
that they are in fact on highly eccentric radial orbits and are just
now executing their first passage. 
Radial orbits are not uncommon in the local group: 
\citet{howley08:ngc205.orbit} recently demonstrated 
that NGC 205 (an M31 dwarf) is also likely on a radial 
passage near its escape velocity; Leo I is similar \citep{mateo:leo1.orbit,
sohn:leo1.kinematics,munoz:leo1.kinematics}. 
This, and the arguments above, 
suggest that constraints regarding the merger history of the 
Milky Way and disk heating should be revisited in light of 
modern cosmological expectations. 

One might think that the difference between the conventional lore on
satellite heating and allowing for more realistic orbits would amount
to a simple numerical correction -- i.e.\ that the differences
outlined above can be subsumed into an appropriate correction of the
efficiency of disk heating, but will not fundamentally alter the
scaling laws derived in the past (namely that disk heating is linear
in merger mass).  In what follows, we show that this is not the case.  At a
basic level, the linear scaling resulting from earlier assumptions is
pathological, and allowing for moderate deviations in line with what
is expected in a $\Lambda$CDM cosmology yields a {\em qualitatively}
different scaling and a {\em non-linear} dependence of disk heating on
satellite mass.  This implies more than an order-of-magnitude
suppression of disk heating in intermediate mass-ratio minor mergers,
and essentially no contribution to disk heating from very minor
mergers (unlike if the heating were linear, in which case there would
be equal contributions to disk thickness from all logarithmic
intervals in satellite and subhalo masses).  We show that the scaling
we derive, non-linear (second-order) in the satellite mass, is
dynamically inevitable and borne out in numerical simulations.

We outline the physics of interest in \S~\ref{sec:why}, explaining why
cosmologically realistic merging orbits (even when nearly circular
initially) tend to be highly radial by the time of coalescence and how
this differs from the assumption of decaying circular orbits
(\S~\ref{sec:orbits}), demonstrating how this changes the qualitative
scaling of disk heating with mass ratio (\S~\ref{sec:heating}),
deriving full analytic expressions for the disk thickness and vertical
velocity dispersion as a function of merger and disk properties for
the radial cases of interest (\S~\ref{sec:equations}), and testing
that these equations accurately describe the results of full numerical
simulations (\S~\ref{sec:tests}).  We discuss the accuracy and
robustness of these approximations in \S~\ref{sec:robustness},
examining when they are or are not appropriate.  In
\S~\ref{sec:consequences} we consider the consequences of the
different scalings owing to radial versus circular orbits, in terms of
the allowed merger histories of the Milky Way and other massive disk
galaxies (\S~\ref{sec:consequences:mergerhistory}) and the shapes of
galaxies along the Hubble sequence (\S~\ref{sec:consequences:shapes}).
Finally, we summarize our results and discuss some of their
cosmological implications and applications to other models and
observations in \S~\ref{sec:discussion}.

\breaker
\section{Why Do Realistic Orbits Make a Difference?}
\label{sec:why}

\citet{tothostriker:disk.heating} developed a simple model for disk
heating by a merger with a satellite.  They assumed that the satellite
-- of mass $M_{2}$, with respect to the primary mass $M_{1}$ -- was on
a circular or nearly circular orbit, which decayed by dynamical
friction. Since the energy of an orbit with semi-major axis $a$ is
$\sim G\,M_{1}\,M_{2}/a$, the orbital energy lost by the satellite
(which they assumed was converted into kinetic energy of the nearby
stars and dark matter, which themselves are the agents slowing down
the orbit) in some annulus as the orbit
decays from $a\rightarrow a-{\rm d}a$ is $\sim
(G\,M_{1}\,M_{2}/a)\,({\rm d}a/a)\propto M_{2}$.  Since the vertical
potential for an isothermal sheet $\propto g\,H$ and the kinetic
energy $\propto \sigma_{z}^{2}$, given the assumption that once the
disk relaxes it will obey the virial theorem ($E = 3\,T$, for a
planar geometry) this implies that the local scale height $H$ is
proportional to the total energy deposited in that annulus ($H\propto
M_{2}$, if the disk was initially cold).  In detail, they obtained that
at a cylindrical radius $R$ in the disk the average change in scale
height owing to a merger is
\begin{equation}
\frac{\Delta H}{R} = {\Bigl(}\frac{0.49}{1+0.434\,\tilde{H}_{0}}{\Bigr)}\,{\Bigl(}\frac{M_{2}}{M_{1}[<r]}{\Bigr)},
\end{equation}
where $\tilde{H}_{0}\equiv 10\,H_{0}/R \sim 1$ is some internal (non-merger induced) scale height, 
but in any case the dependence on this term is weak; the primary scaling is 
${\Delta H}/R \sim M_{2}/M_{1}$. 
Indeed, simulations confirm that this scaling does reasonably 
describe the behavior of the disk in the limit of 
of a rigid satellite on a circular orbit \citep[see e.g.\ the discussion in][]{benson:heating.model}, 
and that adopting the local virial theorem is a good approximation 
for deriving the scale height of the disk, given some energy deposition 
\citep{quinn93.minor.mergers}.

\subsection{How Are Real Orbits Different?}
\label{sec:orbits}

However, a fundamental assumption here, that the orbits are circular
and decay slowly (in a gentle circular in-spiral) by dynamical
friction while the satellite remains intact, bears little resemblance
to realistic orbits in $\Lambda$CDM cosmologies.  It is now known that
satellites merge on highly eccentric, nearly radial orbits \citep[see
e.g.][]{vandenbosch:no.orbital.circularization.due.to.dyn.frict,
benson:cosmo.orbits,khochfar:cosmo.orbits}.  Moreover, the decay of
the orbit, while broadly governed by similar scalings such that e.g.\
the dynamic friction timescale is not a bad approximation \citep[in an
order-of-magnitude sense; see e.g.][]{
vandenbosch:no.orbital.circularization.due.to.dyn.frict,velazquezwhite:disk.heating,
boylankolchin:merger.time.calibration}, is usually nothing like the
smooth circular decay assumed in \citet{tothostriker:disk.heating}. 

Typically, the infalling satellite and primary are deformed after
first passage, and resonances between the internal frequencies of the
systems and their orbital frequencies allow efficient transfer of
angular momentum from the orbit to the halo, while the systems linger
near apocenter after the passage. They then fall in on a nearly radial
orbit and quickly merge, with the braking of the secondary orbit
occurring in a couple of quickly damped radial oscillations through
the center of the primary. Even if first passage is at large
impact parameter, this process will still generally play out, but
perhaps with one or two additional, increasingly radial, passages. At
the same time, once the satellite approaches the radii of interest --
where it is actually crossing and penetrating the disk, so that disk
heating can be effective -- it rapidly begins to be tidally stripped
and destroyed by the gravitational interaction, and cannot execute a
large number of orbits without losing some significant amount of mass
(such that only the first couple of passes are the dominant
perturbation, not the sum over all passages).

The radial nature of these passages in part owes to cosmologically
radial orbits, but it is
also an inevitable consequence of {\em any} orbits that will
actually merge in a dark-matter dominated universe.  To see why this
is true, consider the nature of dynamical friction for a satellite at
some instantaneous three-dimensional radius $r$ from the primary
center.  Technically, the approximation of slow circular orbital decay
owing to dynamical friction, which implicitly relies upon the
background being infinite, homogeneous, isotropic, and (in a bulk
sense) unchanged by the motion and gravity of the secondary, is valid
only when the enclosed mass within a sphere of radius $r$ is much
greater than the gravitational mass (baryons plus whatever dark matter
halo remains bound) of the secondary. So, for example, a small
secondary far out in the halo (e.g.\ the common case of a satellite at
large $r$ in a massive cluster) can experience orbital decay owing to
traditional dynamical friction.

However, once it reaches a radius where the enclosed mass is somewhat
comparable to the satellite mass, this orbital decay approximation is
clearly no longer appropriate (consider e.g.\ the limit of two
identical point masses, one the satellite -- when they are close
enough that the mass of the background halo is negligible, they would
just enter into a stable, marginally bound binary orbit, and coalesce
slowly). Effects like these are well-known even in numerical simulations 
that model the idealized case of a rigid satellite in a smooth halo 
\citep[see e.g.][who argue for a radially-varying Coulomb logarithm 
to match N-body results and the analytic prediction]{hashimoto03:varying.culomb.log.in.dynfric}, 
although these do not include the resonant effects that ultimately 
remove most of the orbital angular momentum. 
For a satellite that is small relative to the primary
($M_{2}/M_{1}\ll0.1$, say), this radius may be negligibly
small. However, for a satellite that is even a moderate fraction of
the primary mass ($M_{2}/M_{1}\gtrsim 0.1$), this will happen, for any
reasonable mass profile, when the secondary gets near of order the
scale length of the primary (considering the halos of the systems,
this is really the scale length of the primary halo -- i.e.\ substantially 
larger even than the scale length of the primary disk). This is precisely
the radius of interest for disk heating. In other words, by the time
the secondary is close enough to the primary that it can actually
move through and heat the stellar disk, it is, by definition (for
any secondary which is a significant mass fraction of the primary) in
a regime where dynamical friction and circular orbital decay are no
longer valid approximations.

Rather, as we noted above, it is in this situation that resonances
between the internal velocity of the satellite and its orbital motion
will allow efficient transfer of orbital angular momentum, leading
to a radial infall.  Because these resonances act efficiently when the
internal velocities of the galaxies are comparable to their orbital
velocities, it is precisely the regime where the secondary is near the
primary -- where circular orbital decay from dynamical friction ceases
to be a valid approximation -- when these mechanisms act
efficiently. Because internal motions (of at least some parts of the
halo) are in resonance with those of the orbit, that dark matter is
strongly accelerated, removing angular momentum from orbital motion
far more efficiently than dynamical friction, allowing an orbit to
transform from nearly circular to nearly radial in just one or two
passages. Since the angular momentum of the orbit is transferred
quickly in a passage (where the system spends most time near some
apocenter), the satellite cannot gently spiral in but must return from
apocenter in a radial plunge. For a satellite falling in on a radial
orbit, it is effectively a free-fall at the escape velocity.  The
braking by dynamical friction on this infall is weak as the system
free-falls (recall, it takes many circular orbits for the orbit to
decay by standard dynamical friction).  

Rather, when the secondary
finally punches through the primary and begins to oscillate through
the primary on a radial orbit, violent relaxation (the scattering of
stars in the secondary and in the center of the primary) off of the
time-dependent central potential transfers energy from the radial
orbit of the secondary to the motion of the central stellar
distribution (preferentially to whatever stars are already ``hottest''
in the center of the galaxy). This quickly damps out the orbit,
completing coalescence in a couple of dynamical times.  The energy of
the orbit, by this rationale, will largely have gone into heating the
bulge of the remnant. This is both because the bulge dominates the 
stellar mass density at the center of the galaxy, and because 
the hot/spherical component of the galaxy (having a more isotropic 
orbit distribution, with a much larger population on radial 
orbits compared to a cold disk) is in resonance with the orbital 
decay of the secondary. 
Some new bulge will be created by violent
relaxation and heating of the central stellar populations -- we show
in \citet{hopkins:disk.survival} that the population violently relaxed
is given by the stars within a radius enclosing a stellar mass
comparable to the stellar mass of the secondary. In short, a mass and
energy $\sim M_{2}$ is added to a classical bulge from the infall, not
to heating the disk.

A similar caveat applies to considering a constant-mass
secondary. Implicit in the circular orbit in-spiral approximation --
in fact, critical to the $\Delta H/R \propto M_{2}/M_{1}$ scaling --
is the idea that the satellite heating is integrated over a large
number of orbits crossing a given radius $r$ while the orbital angular
momentum gradually decays.  Clearly, if the satellite is significantly
stripped of mass and/or partly destroyed by shocking or tidal
interactions after a few orbits, then it is only the first single
passage or couple of passages that dominate the heating. In such a
case, modulo a geometric prefactor, the effects should be identical to
the single radial infall or plunge scenario described above (even if
the orbit is still formally perfectly circular). In short, if the
satellite loses mass after a couple of passages, then the dominant
passages seen by the disk will be the first couple, which will heat
the system the same whether they are circular or radial (the exact
direction in which the secondary is traveling does not enter into the
second-order heating below).

Another way of phrasing this is to note that a significant fraction of
the orbital energy is no longer lost to heating the disk, but carried
away by the tidally stripped material \citep[which could itself heat the
disk in principle, but numerical experiments show that it generally is
added to the halo and in more spherical fashion than the disk, while
retaining its coherent orbital motion rather than heating nearby
material, and moreover because the material is, essentially by
requirement in order to be stripped, thin and diffuse relative to the
potential nearby, it cannot effect significant heating in the manner
of the entire satellite if it moves through the disk; 
see e.g.][]{kazantzidis:mw.merger.hist.sim}.  Moreover, this
process can accelerate the orbital angular momentum loss,
by shedding angular momentum from the orbit preferentially in the
stripped material (there is a general tendency for angular momentum
transport to the least-bound material in interactions). 

As was the case above for the resonant interactions that remove
angular momentum, we can roughly derive the radius where stripping is
expected to become important, and find (unsurprisingly) that it is
again precisely the radius of interest, i.e.\ the radius where the
secondary begins to actually interact with and cross the primary disk
(since the primary disk is much more dense than its extended halo,
this will almost always be the case). Even ignoring the more dense
baryonic components, if we model the secondary and primary as
\citet{hernquist:profile} profile spheres with effective radii
$a\propto M^{1/3}$ \citep[as expected and observed for both halos and
disks; see e.g.][]{momauwhite:disks,shen:size.mass}, then calculate
the formal tidal radius $\tilde{r}=r_{2}/a_{2}$ within the secondary
as a function of the separation $x=r/a_{1}$, the dependence on mass
ratio cancels out entirely and we have $\tilde{r}\,(1+\tilde{r})^{2} =
(1+x)^{3}/2$ -- i.e.\ material within a fraction $\approx
0.8\,(r/a_{1})$ times the effective radius of $M_{2}$ will be stripped
when the satellite is at a distance $r$ relative to the effective
radius $a_{1}$ of the primary. In short, precisely the time of
interest, when the satellite enters the effective radius of the
primary, is when the satellite begins to be rapidly stripped 
\citep[for a more details, see][]{benson:heating.model}.

\subsection{How Much Heating Arises From Realistic Orbits?}
\label{sec:heating}

How much disk heating is effected in this process? There are several
ways to cast this problem, but all yield the same behavior.  For
example, we can even obtain the correct scaling by considering a pure
dynamical friction derivation, despite the caveats above, so long as
we acknowledge the more realistic nature of the orbits.  For
simplicity (ignoring temporarily the caveats just given) let us assume
that the secondary moves through an isotropic homogeneous background
density, with (for analytic convenience, although it makes little
difference to the end scaling) an isothermal sphere profile. The
instantaneous effective dynamical friction force on the secondary is
\begin{equation}
{\bf F}_{\rm df} = -0.428\,\ln{\Lambda}\,\frac{G\,M_{2}^{2}}{r^{2}}\,\hat{v},
\end{equation}
where $\ln{\Lambda}$ is the appropriate Coulomb logarithm \citep[we
will suppress this for now, but including it is comparable to e.g.\
the uncertainties owing to details of the profile shape, and numerical
experiments suggest $\ln{\Lambda}\approx1-2$;
see][]{quinn86:dynfric.on.sats,velazquezwhite:disk.heating,
boylankolchin:merger.time.calibration}.  The force is, by definition,
in the direction of the secondary velocity vector ${\hat{v}}$ (and
opposed to the velocity, hence being negative). If the secondary
travels a distance ${\rm d}{\bf s}$, then, an effective work is done
by dynamical friction involving a transfer of energy from the orbit of
the secondary to the motions of the stars/dark matter that define the
background, with magnitude ${\bf F}_{\rm df}\cdot{\rm d}{\bf s}= |{\bf
F}_{\rm df}|\,|{\rm d}{\bf s}|$.

If the satellite is on a circular orbit, then this motion ${\rm d}{\bf s}$ is 
largely in the azimuthal direction, and the radius $r$ decays slowly. 
The tangential dynamical friction force gives a rate of change of 
specific angular momentum $L$ of ${\rm d}L/{\rm d}t = F_{\rm df}\,r/M_{2}$. 
Since for a circular orbit $L=r\,V_{c}$ (where $V_{c}$ is constant 
for the isothermal sphere), and ${\rm d}s=V_{c}\,{\rm d}t$, 
we can rewrite this as ${\rm d}r = -0.428\,(G\,M_{2}/V_{c}^{2}\,r)\,{\rm d}s$, 
if the satellite remains constant internally throughout the orbit(s). 
Since the enclosed mass within $r$, $M_{\rm enc}=V_{c}^{2}\,r/G$, 
this is ${\rm d}r = -0.428\,(M_{2}/M_{\rm enc})\,{\rm d}s$. 
In order for the circular orbit to decay in radius by ${\rm d}r$, then, 
a work 
$(G\,M_{2}\,M_{\rm enc}/r)\,({\rm d}r/r)$ must be done. 
Since $M_{1}\approx M_{\rm enc}$, we have just recovered the 
\citet{tothostriker:disk.heating} formula, for the specific case of an isothermal 
sphere. The imparted energy in the annulus ${\rm d}r$ is 
$\propto M_{1}\,M_{2}$, so the relative heating 
(i.e.\ imparted energy relative to the characteristic energy of 
stars at $r$ in $M_{1}$, $\sim G\,M_{1}^{2}/r$) 
goes linearly in the mass ratio, $\propto M_{2}/M_{1}$. 
In short, even though the instantaneous force $\propto M_{2}^{2}$ 
as the secondary orbits, it takes $\sim (M_{1}/M_{2})\,{\rm d}r/r$ 
orbits to decay in radius ${\rm d}r$, so the total energy 
imparted in an annulus scales $\propto M_{1}\,M_{2}$ (note that this is 
why it is important that the satellite remain intact in this derivation, 
since we have implicitly assumed that the same mass $M_{2}$ 
makes $\sim M_{1}/M_{2}$ crossings of the disk). 

The situation is trivial for a radial orbit. Here, ${\rm d}s={\rm d}r$, 
so the work done as the secondary falls inwards by ${\rm d}r$ is 
just $0.428\,(G\,M_{2}^{2}/r)\,({\rm d}r/r)$. 
The imparted energy $\propto M_{2}^{2}$, so the relative 
heating is quadratic in the mass ratio, $\propto (M_{2}/M_{1})^{2}$. 
Note that this is valid in an instantaneous sense (i.e.\ with $M_{2}$ measured 
as the system enters the radius $r$) even if the system is being 
stripped or destroyed with each encounter in the disk. 

We could instead treat the problem entirely differently, and obtain
the same answer. For example, assume instead that we face the opposite
regime, where the orbit punches through the disk rapidly
instead of being gently dragged, in which case the impact
approximation is reasonable. Moreover, this could be used to describe
the response of the disk at large $r$ to the damping radial
oscillations as the secondary rapidly merges in the center.  In either
case, a full impact approximation solution \citep[derived
following][]{binneytremaine} follows for circular orbits
and a rigid satellite, assuming the impact parameter $b=r$ and impact
velocity $V_{i}\sim V_{c}$, and allowing for a number of independent
passes $\sim M_{1}/M_{2}$ as required by the energetics for
the circular orbit to decay.
This yields the same result, that the energy
imparted at $r$ goes as $\propto G\,M_{1}\,M_{2}/r$.  For a radial
orbit (or allowing for satellite destruction), 
we instead assume a penetrating encounter (the radial limit
$b=0$), with a rapid decay (technically we can fully solve with some
exponential radial decay time or stripping/destruction time
$\tau\propto t_{\rm dyn}$, but in any case this only is relevant for
the numerical prefactor), and we obtain the same scaling as above,
with the energy imparted $\propto G\,M_{2}^{2}/r$.  The details of our
derivation only change the exact numerical coefficients.

It is easy to show that these scalings more or less have to be true.
For circular orbits and a 
rigid satellite, as described at the beginning of this section, one 
can simply take the difference in energy between a circular orbit 
at radius $r$ and one at radius $r+{\rm d}r$, which is 
$(G\,M_{1}\,M_{2}/r)\,({\rm d}r/r)$ -- this energy must go somewhere, 
so modulo the degree to which it is spread over the disk 
(i.e.\ the Coulomb logarithm), the scaling must be similar. 
For a radial orbit, there is no difference in orbital energy in a single 
infall at $r$ and $r+{\rm d}r$, and even in the circular case 
only the first couple of passages matter, 
so the orbital energy loss is negligible (it takes 
many circulations for the orbit to decay). The energy exchange must 
instead come 
from the deceleration from the wake or interaction of the secondary 
with the extended primary -- i.e.\ the back-reaction of the surrounding 
material from the acceleration imparted by $M_{2}$ itself. 
This effect is necessarily second-order in $M_{2}$ (and arises 
only because the real distributions of material are extended, 
rather than point sources). So long as the radial orbit decays 
quickly once it actually penetrates the galaxy with small impact 
parameter, there will not be some large number of passages 
at large $r$ that could change this (and it is almost impossible to 
arrange a realistic stellar distribution that will not quickly damp a 
penetrating radial orbit once it passes through the center of the 
system).

\subsection{A Re-Derivation of Disk Heating for Realistic Orbits}
\label{sec:equations}

Here, we present some useful equations for disk heating owing to 
more realistic orbits. Most of the details 
are derived in \citet{tothostriker:disk.heating}. 
The key quantity that differs, depending on the nature of the 
orbit, is the energy deposition or exchange in 
an annulus of width ${\rm d}r$. It is convenient to 
express this in terms of the absolute value of 
the energy exchanged between orbit and background 
(i.e.\ the work done $W$, or the energy lost in the orbit) per 
unit area, $\Delta e = {\rm d}W/{\rm d}A$. 
For an isothermal sheet potential (a reasonable approximation in a 
sufficiently local patch of the disk), the potential 
(in the vertical direction) is proportional to the scale height 
$H$ times the local surface gravity and the vertical kinetic energy 
is just ${\rm d}M\,\sigma_{z}^{2}/2$. If the system obeys the local 
virial theorem then $E=T+U=3\,T$, and it is straightforward to calculate 
the heating from some energy input $\Delta e$: 
\begin{eqnarray}
& & H(r) = \frac{2}{3\,\pi} \frac{\Delta e}{G\,\Sigma_{\rm d}(r)^{2}\,(1+\eta_{h}(r))}\\
& & \sigma^{2}(r) =  \frac{2}{3} \frac{\Delta e}{\Sigma_{\rm d}(r)\,(1+\eta^{\prime}_{h}(r))}.
\end{eqnarray}
Here, $\Sigma_{\rm d}(r)$ is the disk surface density at $r$ 
and the term $(1+\eta_{h}(r))$ is an order unity (generally small 
around $R_{\rm d}$, the scale radius of the disk) correction for the disk-halo gravity. 
That these are good approximations to full numerical results 
(e.g.\ that the system obeys the local virial theorem and that 
the isothermal sheet approximation is reasonable) has been 
confirmed in a number of experiments \citep{quinn93.minor.mergers,
walker:disk.fragility.minor.merger,
velazquezwhite:disk.heating,benson:heating.model}. 

Based on our derivation above, we expect for a radial orbit 
moving through a background of density $\rho$ that 
$\Delta e = (F_{\rm df}\,{\rm d}r) / (2\pi\,r\,{\rm d}r)$, giving
\begin{equation}
\Delta e = \alpha\,\frac{G^{2}\,M_{2}^{2}\,\rho}{v^{2}}
\end{equation}
in full generality, where 
$\alpha$ is the appropriate constant depending on the 
Coulomb logarithm and the isotropy of the background velocity 
distribution. 

Up to this point, we have made no distinction between a
dissipationless and dissipational disk. If there is gas in the disk,
it can radiate energy efficiently. For simplicity, assume the gas is
distributed in the same manner as the stars in the disk with gas
fraction $\fgas=M_{\rm gas}/(M_{\rm gas}+M_{\ast})$, and that the
energy correspondingly couples evenly to the gas and stars (i.e.\ a
fraction $f_{\rm gas}$ couples to the gas).  If the gas can
efficiently radiate that energy, then the coupled energy of interest
(that which will not just radiate away) is $\Delta e_{\rm
eff}=(1-\fgas)\,\Delta e$. Given this effective energy injection, our
previous equations for $H(r)$ and $\sigma^{2}$ are unchanged (they
will change only on the 2-body relaxation time of the local stellar
disk, as energy can be re-exchanged between individual stellar orbits
and the thin gas disk, but this timescale is long relative to the
heating of interest).

For simplicity, assume the halo is an isothermal sphere, such that the 
circular velocity is approximately constant as a function of radius. 
This then reduces to 
\begin{equation}
\Delta e = \frac{\alpha}{2\,\pi}\,(1-\fgas)\,\frac{G\,M_{2}^{2}}{r^{3}}
\end{equation}
and 
\begin{eqnarray}
\label{eqn:H}
& & \frac{\Delta H}{R_{e,\,{\rm disk}}} = 
\alpha_{H}\,(1-\fgas)\,{\Bigl(}\frac{M_{2}}{M_{\rm d}}{\Bigr)}^{2}\ \tilde{h}(R/R_{e,\,{\rm disk}}) \\ 
\label{eqn:sigma}
& & \frac{\Delta \sigma_{z}^{2}}{{V_{c,\,{\rm disk}}}^{2}} =
 \alpha_{\sigma}\,(1-\fgas)\,\,{\Bigl(}\frac{M_{2}}{M_{\rm d}}{\Bigr)}^{2}\ \tilde{s}(R/R_{e,\,{\rm disk}})
\end{eqnarray}
where $\alpha_{H}\approx1-2$ and $\alpha_{\sigma}\approx0.3-0.6$ are the appropriate 
normalization constants and 
$\tilde{h}(R/R_{d})\sim \tilde{s}(R/R_{d}) \sim1$ are weak functions of $R/R_{d}$ 
(for convenience, we define them so that $\tilde{h}(1)=\tilde{s}(1)=1$). 

In detail 
these quantities will depend 
on how the orbital decay proceeds as a function of radius. 
For the precise case derived above on a purely radial orbit 
that decays instantly at $r\rightarrow0$ (and isotropic heating), we 
obtain if the disk is a \citet{mestel:disk.profile} disk (constant circular velocity $V_{c,\,{\rm disk}}$, 
where we define $\tilde{v}=(V_{c,\,{\rm disk}}/V_{\rm h})^{2}\approx1$ relative to the 
halo circular velocity, and  
$\Sigma_{\rm d}\propto 1/R$)
$\alpha_{H}=1.60\,\tilde{v}$ 
and $\tilde{h}(x)\approx[0.7\,x+0.3\,\tilde{v}]^{-1}$ \citep[giving a roughly constant $\Delta H$ over 
the radii of interest, similar to what is observed; see e.g.][]{degrijs:scale.heights.vs.r,
bizyaev:scale.heights.vs.r}, 
with $\alpha_{\sigma}=0.57\tilde{v}$, $\tilde{s}(x)=x^{-2}$. 
For an exponential disk (with $V_{c,\,{\rm disk}}$ defined as the maximum circular 
velocity of the disk) we obtain 
$\alpha_{H}=0.68\,\tilde{v}$, $\tilde{h}(x)=[2\,x^{2}\,\exp{(-x)}+0.26\,\tilde{v}]^{-1}$ 
and $\alpha_{\sigma}=0.32$, $\tilde{s}(x)=x^{-2}$. 
% alpha_sigma should be a factor of 2 higher, but its b/c considering sigma_z here -- 
%   so half the energy assumed to go into sigma_R, half into sigma_z

However, more realistically, the 
normalization here reflects a sum over various passages 
as the complex resonant interactions described above rapidly 
remove angular momentum from the secondary. At the same time, 
tidal stripping is continuously removing mass, and the gravitational 
mass of the primary can effectively appear different depending 
on where in the potential the secondary is located. 
It is possible to develop a much more complex analytic model that 
attempts to account for these effects, including e.g.\ some model of 
dynamical friction, stripping, disk-satellite shocking, and then 
time-integrating these equations to produce the numerical constant 
desired \citep[see e.g.][]{benson:heating.model}. Although these offer 
some improvement, it is still unfortunately not possible to follow the proper 
sources of angular momentum loss (resonant coupling in the orbit) 
and back-reaction of the halo and secondary in fully analytic form, 
and because of this, these more sophisticated treatments do not 
yield a 
major gain over our much simpler scalings above. 

Fortunately, testing these predictions in full numerical simulations 
(see \S~\ref{sec:tests} below) -- 
including live halos and secondaries, systems with a huge number of 
very small satellites, and gas-rich disks evolved in full hydrodynamics -- 
we find that the scalings above nevertheless accurately 
describe the disk response, with a simulation-to-simulation scatter of 
only a factor $\sim2-3$. Calibrating the normalization of the 
efficiencies properly to N-body results, we obtain 
$\alpha_{H}\approx1.0-2.0$ and $\alpha_{\sigma}\approx0.3-0.6$, 
similar to the coefficients estimated analytically. We find that 
these calibrations are robust across a wide range of simulations 
and mass ratios, and describe the behavior of $\Delta H/R$ and 
$\Delta\sigma^{2}/v_{c}^{2}$ well from 
$\sim 1-5\,R_{d}$ (outside of which the disk is fairly ill-defined, 
both numerically and physically, so we are not surprised that the 
scalings do not trivially extrapolate to infinite $R$).

\subsection{Numerical Tests}
\label{sec:tests}

In order to test how robust these scalings are, we compare them 
with N-body simulations. 
We compile results from a number of different systematic studies of 
mergers, in order to see whether the particular choices in a 
given study are pathological. 

\citet{velazquezwhite:disk.heating} consider 
collisionless mergers of an N-body disk+bulge+halo 
Milky Way-like model 
primary (although the system is collisionless, and so 
can be freely rescaled to arbitrary mass) 
with N-body \citet{king:profile}-model (truncated isothermal 
sphere) secondary satellites of varying concentration 
and mass ratios 1:5 and 1:10. The authors perform a
survey of orbital inclination (from prograde $\theta=0^\circ$ 
to retrograde $\theta=180^\circ$) and impact 
parameter (equivalent to varying the initial eccentricity or 
circularity of the orbit, with $\epsilon_{J}\equiv J/J_{C}(E)$ 
being the ratio of the orbital angular momentum $J$ to that 
of a circular orbit $J_{C}$ varied). 

\citet{villaloboshelmi:minor.mergers} consider dissipationless 
mergers of a $z=0$ Milky Way-like primary 
and a scaled $z=1$ analog, with disk and satellite properties 
scaled according to theoretical expectations following 
\citet{momauwhite:disks} \citep[note that this may in fact over-estimate the 
redshift corrections, making their $z=1$ case more appropriate 
for $z\sim2-3$; see][]{somerville:disk.size.evol}. They 
model the primary halo as adiabatically contracted, and 
merge it with either bulge+halo or disk+halo satellite 
models of mass ratio 1:5, on varying orbits with a moderately 
radial impact parameter.

\citet{younger:minor.mergers} consider full hydrodynamic merger
simulations of two bulge+disk+halo systems, with mass ratios 1:8, 1:4,
1:3, and 1:2. In each case, the orbital inclination is surveyed, and
two sets of models are run: one with a relatively low gas fraction
$\sim0.2$ in the disk at the time of the merger (corresponding to
Milky Way like disks over the last few Gyr) and one with intermediate
gas fractions $\sim0.4$ (corresponding better to intermediate redshift
$z\gtrsim1$ cases).  Most of the orbits are parabolic and radial
($b\sim1-2\,R_{d}$), but the authors consider limited surveys in the
impact parameter and corresponding circularity/initial angular
momentum of the orbit (from $b=0-10\,R_{d}$). Because, for
dissipational systems, the gas properties set an absolute units scale,
the physical mass and size scale of the galaxies cannot be freely
rescaled: the authors examine primarily a Milky Way-like primary, but
also consider a limited mass survey spanning a factor $\sim10$ in
lower primary masses.

\citet{hayashichiba:subhalo.mergers} investigate dissipationless 
mergers of a disk+bulge+halo primary (including both rigid halos+bulges 
and restricted halos free to move but not realized in N-body fashion, 
with an N-body disk; note that \citet{velazquezwhite:disk.heating} 
argue that these restricted 
halo models may artificially inflate the resulting scale heights by a 
factor $\sim2$, but this is comparable to the scatter in any case). 
The halo is merged with a large number of satellites (subhalos), drawn 
in Monte Carlo fashion from a 
mass spectrum with power-law index $dN/dM\propto M^{-2}$, 
similar to the sub-Milky Way end of the halo mass function 
\citep[see also][who perform a similar exercise and reach similar 
conclusions]{ardi:subhalo.mergers}. 
A mass fraction $=0.1$ of the primary is placed into satellites 
according to that mass distribution, with a their spatial distribution 
varied systematically (following a \citet{hernquist:profile} profile), 
and their initial velocity distribution 
determined assuming they follow the local halo velocity ellipsoid 
with either an isotropic dispersion tensor 
($\beta=1-0.5\,(\sigma_{\theta}^{2}+\sigma_{\phi}^{2})/\sigma_{r}^{2}=0$)
or a radially anisotropic dispersion ($\beta=0.5$). In 
addition to their spatial and orbital velocity distributions, the absolute 
number of subhalos and maximum and minimum masses sampled 
are varied, and both point-mass subhalos and extended-mass 
(N-body) subhalo models are considered. 

\begin{figure}
    \centering
    \scaleup
    %\plotone{numerical_check.ps}
    \plotterr{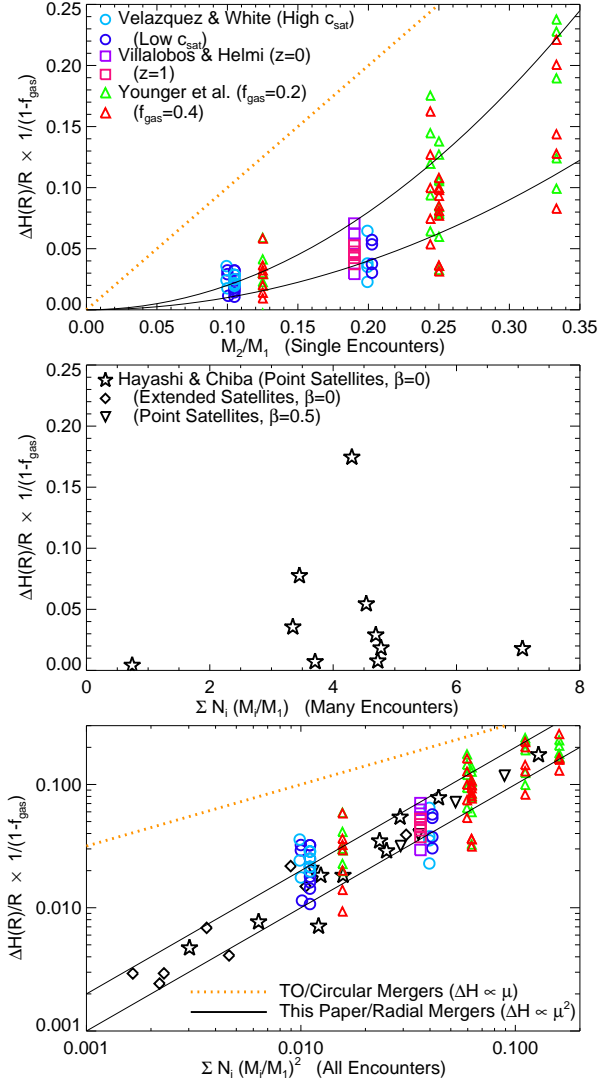}
    \caption{Analytic predictions for disk heating 
    in mergers and results of 
    numerical experiments. {\em Top:} 
    Disk heating (change in scale height at $R\approx R_{e}$) 
    $\Delta H(R)/R$ versus merger mass ratio $M_{2}/M_{1}$. 
    Contrast our prediction $\Delta H \propto (M_{2}/M_{1})^{2}$ 
    (Equation~(\ref{eqn:H}); black lines show a 
    factor 2 scatter) 
    and that of \citet{tothostriker:disk.heating} for  
    rigid satellites
    on circular orbits ($\Delta H \propto M_{2}/M_{1}$; orange dotted). 
    We compare N-body simulations of disk-satellite mergers 
    of various masses (labeled). 
    Each simulation set surveys orbital parameters (inclination,  
    impact parameter, angular momentum). 
    The dissipationless \citet{velazquezwhite:disk.heating} 
    simulations also survey satellite mass profiles 
    and concentrations ($c$); the dissipationless 
    \citet{villaloboshelmi:minor.mergers} simulations scale 
    systems to $z=0$ and $z=1$, and consider 
    both disk and spheroid satellites. The hydrodynamic 
    \citet{younger:minor.mergers} simulations survey 
    disk gas content (disk heating 
    scales $\propto(1-\fgas)$; to factor out this dependence, 
    we plot these points as $\Delta H/R \times 1/(1-\fgas)$).
    {\em Middle:} Heating versus the linear sum over encounters 
    in \citet{hayashichiba:subhalo.mergers} 
    simulations. They 
    consider simultaneous mergers with many ($\sim100-1000$) 
    small satellites ($M_{2}/M_{1}\sim0.001-0.1$). Each point is 
    the net heating, integrated over all encounters, for different initial 
    configurations.  
    If heating were linear in $M_{2}/M_{1}$, there should be a linear correlation 
    between the net heating and the linear sum over mergers, 
    $\Sigma N_{i}\,(M_{i}/M_{2})$. 
    {\em Bottom:} Heating versus the quadratic sum over all 
    encounters, as predicted by our Equation~(\ref{eqn:H}). 
    Numerical experiments confirm 
    our predicted scaling with factor $\sim2-3$ scatter, and demonstrate that the 
    scaling of disk heating is second-order 
    in $M_{2}/M_{1}$. 
    \label{fig:nbody}}
\end{figure}

Figure~\ref{fig:nbody} plots the results of these experiments. 
Specifically, we compile (for the experiments described) the 
resulting scale height of the disk after the merger, 
at some radii $\sim 1-4\,R_{d}$ (i.e.\ within a factor 
of $2$ of $R_{e}$, where we do not expect a strong 
dependence of $H$ on $R$). In all cases, the authors evolve 
corresponding isolated disks, so the value we consider is 
$\Delta H$, the change in scale height relative to an 
isolated disk (i.e.\ subtracting the effects of internal heating). 
We convert all the results to a uniform definition in order to 
compare them in a fair manner: we define 
$\Delta H$ precisely as the median scale-height of stars 
in the annulus ${\rm d}R$ (i.e.\ exactly half the disk stars in the 
cylindrical radial annulus ${\rm d}R$ should 
lie in $z =\pm H$), and obtain $\Delta H$ as close to the 
disk half-mass radius $R_{e}$ as possible 
(but in any case rescale all the results to $\Delta H(R)/R$ 
for the $R$ measured). 

First, we consider the simulations from 
\citet{velazquezwhite:disk.heating}, \citet{villaloboshelmi:minor.mergers}, and 
\citet{younger:minor.mergers}, who consider idealized single 
encounters (as opposed to large numbers of multiple or simultaneous encounters). 
We compare $\Delta H(R)/R$ as a function of the mass ratio of the 
encounter $\mu=M_{2}/M_{1}$, for the various simulations, to the analytic 
predicted scalings from \citet{tothostriker:disk.heating} (linear in $\mu$) and our derivation herein 
(second-order in $\mu$). 
For the \citet{younger:minor.mergers} hydrodynamic simulations, we 
confirm the predicted scaling 
$\Delta H\propto (1-\fgas)$ \citep[and a more limited subset 
shows that this applies reasonably well over a systematic 
study varying $\fgas=0-1$; see e.g.][]{hopkins:disk.survival}. 
To compare uniformly with the other 
(dissipationless, $\fgas=0$) 
numerical simulations here, we actually plot $\Delta H/R \times 1/(1-\fgas)$, 
so that all simulations are on the same footing. 
For each set of simulations and mass ratio, there is a factor $\sim2-3$ scatter 
owing to a combination of orbital parameters (both orbital inclination and 
impact parameter) and the detailed 
choices for the structural properties of the galaxies (recall, the simulations 
plotted make various and in some cases quite different assumptions 
about the shapes of secondary galaxies and their mass profiles). 
In any case, the scatter is still reasonably small, and despite the fact 
that these simulations include e.g.\ the more complex realistic 
orbital structure, tidal stripping, and other effects, they agree quite well 
with our predicted scalings for the range 
in normalizations $\alpha_{H}\approx1-2$. 

If we were to consider arbitrary normalizations $\alpha_{H}$ (i.e.\
ignore the analytic expectations entirely), then for any single
encounter, there is of course a formal degeneracy between the
dependence of resulting scale height on mass ratio and the efficiency 
(implicit in $\alpha_{H}$).
Even for the set of simulations just described, the dynamic range in
$\mu$ is relatively small and scatter significant, so if we freely
renormalized the \citet{tothostriker:disk.heating} prediction, we
could find a solution that was not a particularly bad match to the
numerical results (albeit still not as good a fit 
as the predicted quadratic dependence, even with a fitted
$\alpha_{H}$ that differs from the analytic expectation by almost an
order of magnitude). The
simulations of \citet{hayashichiba:subhalo.mergers}, however,
include the effects of a large number of encounters with a wide
range of mass ratios ($\sim 2$ orders of magnitude in $\mu$), 
and can strictly break this degeneracy.

If the true scaling of $\Delta H/R$ were linear in $\mu$, regardless
of its normalization, then, in a sum of $N_{i}$ encounters with
various systems of mass ratios $\mu_{i}$, the net heating should be
linear in the sum of these individual events (modulo some scatter),
i.e.\ $\Delta H/R \propto \Sigma\, N_{i}\,\mu_{i}$.
\citet{hayashichiba:subhalo.mergers} consider this, and we show their
results (renormalized appropriately for our definitions) in
Figure~\ref{fig:nbody}. Note that because some of the orbits they
consider are very long axis and/or energetic, a given satellite
(especially in the cases where they use point mass satellites, since
there is no stripping/destruction of the satellite) can make many
well-separated passes through the disk. The appropriate sum is then
over each encounter, treating them separately with $\mu_{i}$
appropriate for the mass ratio at each encounter that penetrates the
disk (for this reason, with some satellites making many passages, the
sum $\Sigma\, N_{i}\,\mu_{i}$ can be $>1$, despite the fact that the
total subhalo mass in these simulations is only a fraction of the
primary mass). Each point in the figure represents the sum over many
small mass ratio encounters with one of the initial configurations
described above, typically mergers with $\sim1000$ satellites spanning
a mass spectrum from $\mu_{i}\sim 10^{-3}-10^{-1}$. In any case, the
result is essentially a scatter plot, with no significant correlation
visible.

If, instead, the appropriate scaling goes as $\mu^{2}$, as we have
predicted, then the net heating in this case should behave as $\propto
\Sigma\, N_{i}\,\mu_{i}^{2}$.  Considering the results from
\citet{hayashichiba:subhalo.mergers} in this comparison, we do indeed
find a tight linear correlation (as the authors themselves note). 
In short, the heating from each of
the many encounters $\mu_{i}$ is clearly adding quadratically -- i.e.\
the effective heating per event scales $\propto\mu_{i}^{2}$, not
linearly. Because this is a mix of mass ratios, projecting versus
$\Sigma\, N_{i}\,\mu_{i}$ degrades the correlation increasingly the more
the merger history is varied.  We also reproduce the simpler merger
history simulations from \citet{velazquezwhite:disk.heating},
\citet{villaloboshelmi:minor.mergers}, and
\citet{younger:minor.mergers} in this plot, with our analytic
expectations.  For these single encounters, we take $N_{i}=1$ with
$\mu_{i}$ being the initial mass ratio, appropriate for the fairly
idealized mergers these represent (although they vary other properties
not considered in the \citet{hayashichiba:subhalo.mergers}
simulations), a reasonable approximation also for the reasons below.

\subsection{How Robust Is This Scaling?}
\label{sec:robustness}

Given the complexity of the full time-dependent behavior in these
simulations, and the wide variation in the physics included and
assumptions made by the different authors, it is remarkable that the
agreement is as good as it is in Figure~\ref{fig:nbody}. Why should
our simple scalings (Equations~\ref{eqn:H}-\ref{eqn:sigma}) describe
the full N-body results as well as they do, with only a scatter of a
factor $\sim2-3$ in $\Delta H/R$ and no dramatic systematic offset
between different simulations?  There are in fact several reasons
for this.

Most important, despite the fact that there is considerable complexity
in the orbital decay of satellites, is that once
it begins to experience close passages and penetrating encounters with
the disk, the satellite is efficiently tidally stripped and 
destroyed within a
couple of passages \citep[see e.g.][]{quinn93.minor.mergers,
velazquezwhite:disk.heating,benson:heating.model}.  
The result is that the approximation of a single ``plunge'' with 
instantaneous decay/destruction is quite good. 
Furthermore, the tendency of resonant interactions 
between disk, halo, and substructure to efficiently remove angular
momentum also serves to make final passages more radial, removing some
of the dependence that might otherwise result from initially different
orbital configurations.  Critically, all of these modern simulations
include live disks and massive dark matter halos. 
In addition, differences between the effects of
individual passages are averaged out in realistic cosmological ensembles
with many encounters. 

These numerical experiments also allow us to ask the question of how
circular or radial a given orbit should be for either the
\citet{tothostriker:disk.heating} derivation or ours, respectively, to
apply. Clearly, given the variation in orbital parameters of the cases
shown in Figure~\ref{fig:nbody} and agreement of the results with our
predicted quadratic dependence on $\mu$, the radial case derived here,
where the imparted energy $\propto (M_{2}/M_{1})^{2}$, is general, and
the circular case in \citet{tothostriker:disk.heating}, with imparted
energy $\propto (M_{2}/M_{1})$, is pathological.  The encounters in
\citet{velazquezwhite:disk.heating} span a range in circularity from
$J/J_{C}(E)=0.3-0.8$, and the impact parameter study from
\citet{younger:minor.mergers} covers $\sim0.1-3$ in the ratio of the
angular momentum to that of a circular orbit. Moreover,
\citet{hayashichiba:subhalo.mergers} and \citet{ardi:subhalo.mergers}
study full orbit distributions, both isotropic and moderately radially
biased, which should include large populations of initially more
tangential orbits. In all cases, the end results still scale more
generally with the predicted quadratic $\propto (M_{2}/M_{1})^{2}$
rather than linear $\propto (M_{2}/M_{1})$.

It is easy to see why this must be the case. Recall, the 
instantaneous work done by the secondary 
as it moves some distance ${\rm d}s$ is $\propto M_{2}^{2}\,{\rm d}s$ 
(second-order in $M_{2}$). 
The case of rigid satellites on circular orbits yields
a linear scaling in $M_{2}$ 
because the number of orbits 
required precisely cancels one order in $M_{2}$ -- i.e.\ in order for the 
orbit to decay inward by ${\rm d}r$, 
it must execute $\sim (M_{1}/M_{2})\,({\rm d}r/r)$ orbits. So 
the net heating as it moves through 
the annulus is the combination of these two factors: 
$M_{2}^{2}\times (M_{1}/M_{2}) \sim M_{1}\,M_{2}$. 
This requires a rather precise cancellation -- if 
any effect can upset this -- i.e.\ if for some reason the large number of 
circulations is suppressed, and in fact the satellite does not 
execute $\sim (M_{1}/M_{2})$ identical orbits for each 
unit radius it decays -- then the scaling will just go 
as $M_{2}^{2}$. 

Such a suppression is almost inevitable: first, 
as discussed above, a realistic extended secondary will rapidly
lose mass and eventually be tidally destroyed as it executes such
circulations. In practice, the secondary, especially in e.g.\ a 1:10
case where it would require $\sim10$ full orbits through each radial
interval to yield a linear scaling, never survives more than a couple
of penetrating passages without losing a significant fraction of its
mass. The subsequent passages are exponentially suppressed as the
secondary loses mass, and a proper re-derivation of the heating 
gives an
energy transfer rate that goes like $M_{2}^{2}[1 +
\exp{(-M_{1}/M_{2})}\,(M_{1}/M_{2})]$ -- so the leading term is similar to
our estimate for a single passage. In terms of the instantaneous
heating, it does not matter whether the orbit is radial or
circular -- our derivation is really more broadly applicable to all
cases where a very long
(secondary mass-conserving) in-spiral is not possible.

Second, as discussed in \S~\ref{sec:orbits}, 
the system does not, in fact, need $\sim M_{1}/M_{2}$ circulations to lose a 
significant fraction of its angular momentum; 
N-body simulations including extended systems typically find 
that angular momentum exchange in resonances (and mass loss) makes 
only one or two close passages sufficient. 

Third, in systems with a large number of bodies (halo 
substructure), there tends to be efficient transfer and
segregation of angular momentum.  The satellites on circular orbits
will preferentially occupy relatively large impact-parameter orbits in
the outer regions of the halo, not coming close to the disk \citep[see
e.g.][]{font:sats.on.large.orbits}. The systems that will have
encounters with the disk are on radial orbits, and so can 
be considered a series of independent highly radial
passages, rather than a continuous circular heating.  It is
difficult to obtain circular orbits at small radii, and usually
involves invoking slow tangential decay of the initially
circular orbits at larger radii owing to dynamical friction
(a slow process, with characteristic timescales of order the Hubble time).

\breaker
\section{Consequences of these Orbits}
\label{sec:consequences}

Having established that disk heating owing to mergers with 
somewhat radial orbits or 
efficient stripping is second-order in $M_{2}/M_{1}$, 
whereas heating owing to mergers with rigid satellites 
on circular orbits is first-order in $M_{2}/M_{1}$, 
what are the consequences of this? 

\subsection{The ``Allowed'' Merger History of the Milky Way and Massive Disks}
\label{sec:consequences:mergerhistory}

First, consider what this means for the allowed merger 
history of a given galaxy. Modulo the numerical prefactor, 
a rigid satellite on a
circular orbit of mass ratio $\mu\equiv M_{2}/M_{1}$ 
will heat the disk to $H/R\sim\mu$. 
For a 1:10 merger on a circular orbit, for example, this translates to
$H/R\sim0.1$, comparable to values observed, setting a relatively 
stringent limit on how many mergers of even this small a mass 
ratio could have occurred without overheating or 
destroying a massive (Milky Way-like) stellar disk. 
However, if (more realistically) the same merger were to occur 
on a radial orbit and/or stripping were efficient, the heating would be suppressed by an 
additional factor $\sim \mu$, giving $H/R\sim\mu^{2}$. 
For the 1:10 merger, then, this is an order-of-magnitude 
suppression in the disk heating! 
The previous 
constraints in fact would allow quite a large number of 
high mass-ratio (or even a few fairly low mass-ratio) mergers 
without problems for disk survival. 

\begin{figure*}
    \centering
    \scaleup
    %\plotone{mw_constraints.ps}
    \plotone{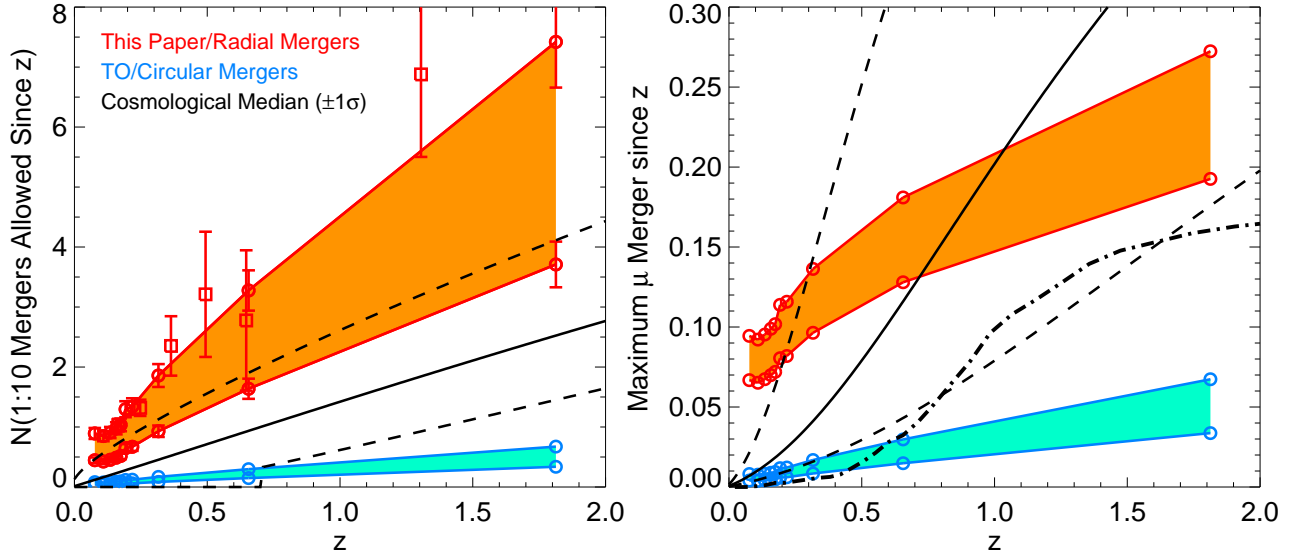}
    \caption{Constraints on the Milky Way merger history from the observed thick 
    disk and local stellar age-velocity dispersion relation. 
    {\em Left:} 
    Number of 1:10 mergers implied by the observed 
    age-velocity dispersion relation, since a given redshift $z$. 
    Each point represents the constraint from observed stellar velocity dispersions 
    in the solar neighborhood \citep[from][circles and squares, respectively]{nordstrom04:mw.age.sigma,
    seabroke.gilmore.07:mw.age.sigma}, at a given stellar age (here converted to a 
    lookback redshift $z$). We convert $\sigma^{2}({\rm age})$ to $N_{\rm mergers}$ 
    assuming all heating comes from gas-free 1:10 mergers 
    (given our analytic Equation~(\ref{eqn:sigma})). Allowing gas makes 
    the constraints more generous by a factor $1/(1-\fgas)$. 
    We contrast $N_{\rm mergers}$ given our derivation of heating in realistic 
    orbits (where $\Delta\sigma^{2}\propto (M_{2}/M_{1})^{2}$; {\em red}) 
    and the derivation from \citet{tothostriker:disk.heating} for rigid
    satellites on
    circular orbits ($\Delta\sigma^{2}\propto M_{2}/M_{1}$; {\em blue}). 
    Shaded range represents the typical factor $\sim2$ scatter in normalization 
    (see Figure~\ref{fig:nbody}). We compare 
    the median $N_{\rm mergers}$ more major than 1:10 since redshift $z$ 
    (solid black line; interquartile range shown as dashed lines) 
    expected for a Milky Way-mass halo in $\Lambda$CDM
    cosmological simulations \citep{fakhouri:halo.merger.rates}. 
    {\em Right:} 
    Same, but instead assuming that all the heating for each point 
    was effected by a single merger -- i.e.\ estimating the maximum allowed 
    merger since $z$. Solid line is the cosmological expectation as before, 
    where halo mass is defined at each redshift; dot-dashed line 
    from \citet{stewart:mw.minor.accretion} is the expectation where the mass ratio is 
    relative to the $z=0$ Milky Way mass (i.e.\ allowing for 
    growth of the disk since a given redshift). 
    Observational constraints, re-analyzed with the appropriate 
    derivations for realistic orbits, easily accommodate the expected 
    merger histories of Milky Way-like galaxies (allowing 
    $\sim5-10$ individual 1:10 mergers).  
    \label{fig:mw.constraint}}
\end{figure*}

Figure~\ref{fig:mw.constraint} re-examines the 
constraints on the Milky Way disk in light of this 
important distinction. Various observations 
\citep{wielen77:mw.age.sigma,carlberg85:mw.age.sigma,
quillengarnett01:mw.age.sigma,nordstrom04:mw.age.sigma,
seabroke.gilmore.07:mw.age.sigma} 
have quantified the vertical velocity dispersion of 
stars in the solar neighborhood as a function of 
the mean stellar population age $\tau$. 
Given the circular velocity $V_{c}\sim 220\,{\rm km\,s^{-1}}$ 
and one of the models for disk heating described in 
\S~\ref{sec:equations} above, this can be translated to an
upper limit on the allowed merger history. Note that some of the observed 
dispersion could owe to purely internal effects -- e.g.\ heating 
by two-body relaxation and scattering off of molecular 
clouds, spiral structure, and other local density perturbations -- 
so assuming that the heating comes entirely from mergers 
as given by the equations above is an upper 
limit to the allowed merger history (technically gas complicates 
this comparison somewhat, since it can dissipate, but as we 
discuss below, the effect is not large and further supports 
our argument that the merger history could be much more 
violent than has been previously assumed). 

In detail, merger histories could be complex and mix many different
mass ratios -- however, in order to show the qualitative result in
illustrative fashion, we convert each value $\sigma(\tau)$, compiled
from the observations of \citet{nordstrom04:mw.age.sigma}, to a
maximal number of 1:10 mergers assuming each merger is exactly
1:10, that there has been no new accretion in the meantime (again,
including such a term would allow even more violent merger histories),
and that each merger heats the disk according to
Equation~(\ref{eqn:H}).  We plot the resulting ``maximum number of 1:10
mergers'' as a function of lookback redshift corresponding to each
stellar age $\tau$, assuming either radial/stripped orbits or rigid satellites 
on circular orbits
in our conversion. We compare this to the median (and $\pm1\,\sigma$
range) number of mergers of this mass ratio or lower (more massive)
integrated from $z=0$ to the corresponding lookback time $\tau$
estimated from the dark matter merger rates as a function of mass and
redshift, for halo masses similar to the Milky Way
($\sim10^{12}\,\msun$, although the expectation is only weakly
mass-dependent), from \citet{stewart:mw.minor.accretion} 
and \citet{fakhouri:halo.merger.rates} (note that for a given definition of merger 
mass ratio, the two calculations give answers in agreement to within $\sim20\%$). 

The answer is as expected from above: if the merging orbits were all circular 
and satellites were rigid, 
transfer of orbital energy to disk heating would have been efficient, and the 
limits are quite stringent (significantly below the expected number of 
mergers). On the other hand, if the mergers were radial (i.e.\ most of their energy 
went into violently relaxing and heating only the central bulge stars) 
and/or stripping acts efficiently after a couple of passages, 
then a much larger number of mergers are permitted -- in fact significantly 
more than are cosmologically expected in typical systems 
(allowing some of the observed heating to stem from 
internal processes, and reflecting the fact that some -- albeit not most -- orbits 
will be more circular than others and yield somewhat more heating). 

Similarly, we can convert the observational constraints into a maximum 
allowed mass ratio merger for each stellar population age (assuming the 
entire heating owed to a single merger, and using Equation~(\ref{eqn:H})
to estimate the corresponding mass ratio for either a circular/rigid or radial/non-rigid 
interaction). 
Figure~\ref{fig:mw.constraint} shows these results as well, compared to the median 
expected maximum 
mass ratio merger since lookback time $\tau$ (and $\pm1\,\sigma$ range). 
The dramatic qualitative difference between orbits 
is again present. Regardless of how we represent merger histories, the 
constraints on disk heating/thickness 
easily allow for the expected cosmological merging histories if the orbits 
are primarily radial and/or 
satellites are non-rigid (with enough cushion that even 
if there is internal heating and the occasional circular orbit, there 
is still no significant tension). 

Note that while our derivation more than allows for the
fraction of $>$\,1:10 mergers in total, there is an intersection of our
constraints with the {\em maximum} mass ratio merger around mass
ratios $\sim$\,1:4-1:3. This, of course, is expected. Roughly half of
the observed galaxy population with the same mass as the Milky Way are
bulge-dominated (S0 or elliptical) galaxies, with most of this mass
assembled since $z\sim1-2$ \citep[see e.g.][]{bell:mfs, borch:mfs}. We
would expect that the Milky Way, being still disk-dominated, would
therefore be below the median (in short, the half of galaxies that
have had a $\sim$\,1:3 merger since $z\sim2$ probably correspond to the
half of galaxies at these masses that are ellipticals). Allowing for
this fact (or equivalently comparing the distribution of expected
mergers in systems that would still be disks today), there is no
tension here (the allowed history for the Milky Way is consistent with
that expected for fully $\sim1/3$ of galaxies of similar mass, or
roughly $\sim 80\%$ of the systems that will still be disk-dominated
at $z=0$). Furthermore, the constraints are even less stringent if we
allow for gas accretion and/or subsequent growth of the galactic disk
after early redshifts.

This is quite generally applicable -- not just to the Milky Way disk, but to 
the population of observed edge-on disk galaxies. For example, 
\citet{dalcanton:age.height.relations} 
estimate the age-scale height relations for a number of observed 
edge-on disks of various masses. The median trends are  
similar to those implied by the Milky Way age-velocity dispersion 
relation
(and in fact if we were to convert their estimates in the same manner in 
Figure~\ref{fig:mw.constraint}, the resulting allowed ranges would lie within 
the same shaded range as estimated for the Milky Way, albeit with larger 
error bars).

\subsection{The Edge-On Shapes of Galaxies Along the Hubble Sequence}
\label{sec:consequences:shapes}

Since both disk heating and bulge formation proceed from 
mergers, the differences above naturally predict a different relation between 
e.g.\ disk thickness and bulge-to-disk ratio along the Hubble sequence 
from very late type disks to bulge-dominated systems. 
In \citet{hopkins:disk.survival} we demonstrate that a mean 
relationship holds between the amount of classical bulge mass 
formed (violently relaxed and formed in a dissipational starburst owing 
to loss of angular momentum in the cold gaseous disks) 
and mass ratio, orbital parameters, and pre-merger gas content 
of merging galaxies. For otherwise fixed parameters at the time 
of merger, the amount of bulge formed scales roughly 
linearly in the mass ratio $\sim \mu$, for the reasons briefly 
discussed in \S~\ref{sec:orbits} above. Correspondingly, the amount of 
disk heating scales according to our estimates, as a function of 
merger mass ratio and gas content. 

A proper quantitative 
calculation of e.g.\ the relation between bulge-to-disk ratios and disk 
scale heights must account for the details mentioned above -- including the 
evolution of merger history as a function of mass ratio and the corresponding 
evolution in gas fractions (which affect both bulge-to-disk ratios and 
disk scale heights, but in somewhat different manners). Moreover, 
these derivations apply only to systems in a post-merger status -- 
more properly we would need to include the effects of accretion of new gas 
subsequent to each event, building up disk mass and making the 
disk more thin. 
Any quantitative estimates therefore require a much more complete 
cosmological model than we have here. 

\begin{figure*}
    \centering
    \scaleup
    %\plotone{isophotes.ps}
    \plotone{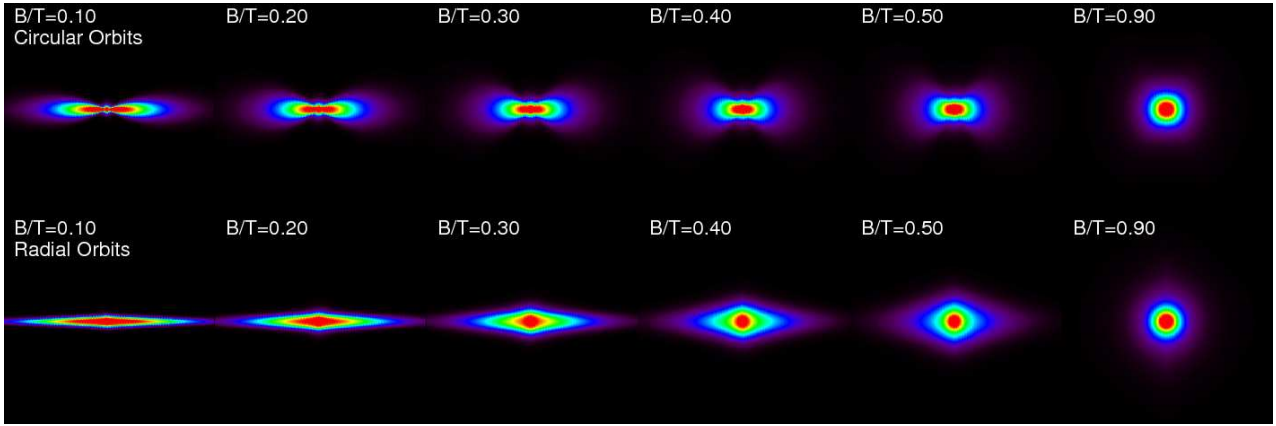}
    \caption{Qualitative (toy model) expectations for the difference in edge-on 
    morphologies of disk+bulge systems (with the given bulge-to-total stellar 
    mass ratio $B/T$) given disk heating according to our derivation for 
    realistic orbits ({\em bottom}) and that for rigid satellites 
    on circular obits ({\em top}). In the toy model, the bulge fraction traces 
    the mass ratio of the most recent (assumed gas-poor) merger, 
    according to the calibration from numerical experiments 
    in \citet{hopkins:disk.survival}; we then calculate the amount of disk 
    heating given this merger, and plot the combined, projected edge-on 
    surface brightness profile of the system. 
    Because disk heating is much more 
    efficient in the rigid satellite, circular orbit regime, disks in 
    intermediate $B/T$ systems are much thicker than observed -- 
    if heights scaled as predicted they would 
    ``swamp'' the bulge and give rise not to a bulge plus disk morphology 
    but to a series of decreasingly elliptical ``boxy'' (rectangular) isophotes. 
    In contrast, our prediction for relatively inefficient heating means the 
    disks even in intermediate $B/T$ systems can be relatively 
    thin ($H/R\lesssim0.2$), leading to a standard 
    bulge plus disk edge-on morphology and 
    the characteristic observed ``disky'' isophotes 
    in large bulges with embedded disks. 
    \label{fig:jz.distrib}}
\end{figure*}

However, we can highlight some of the salient qualitative behaviors
that differ between our derived disk heating expectations for
realistic orbits and the derivation of
\citet{tothostriker:disk.heating}, by adopting a simplified toy model.
Consider a set of disks where the bulge was entirely formed by the
most recent merger, of mass ratio $\mu$, and ignore subsequent
accretion of new disks. The bulge mass fraction will then be $\approx
\mu$, where in detail we use the formulae from
\citet{hopkins:disk.survival} for a median expected orbit (averaging
over random orbits) and small pre-merger disk gas fraction
($\lesssim10-20\%$). 

The surviving disk will have a thickness
determined by Equations~(\ref{eqn:H})-(\ref{eqn:sigma}), or the
corresponding equations from \citet{tothostriker:disk.heating} in the
strict limit of rigid 
satellites on circular orbits -- again we assume $\fgas\ll1$ for
convenience and use the exact normalizations from
\S~\ref{sec:equations}, which agree reasonably well with full
numerical experiments. We assume a disk height $H/R_{d}\approx0.1$ for
isolated systems, comparable to the expectations for purely internal
heating \citep[see e.g.][]{binneytremaine,
sellwood:resonant.disk.thickening} and isolated disk galaxy
simulations
\citep{villaloboshelmi:minor.mergers,younger:minor.mergers}.  These
details are important only for the exact numerical quantities
in the profiles -- the behavior of interest is
that, for the cases we derive here, we roughly obtain $H/R \sim
(B/T)^{2}$ (plus some constant; since $\Delta H/R\propto \mu^{2}$ and
$B/T\propto \mu$), whereas for the \citet{tothostriker:disk.heating}
case we have $H/R\sim(B/T)$ (both $\propto \mu$). The qualitative
relationship between $H/R$ and $B/T$ is different.

Assuming the initial disk is Milky Way-like (giving it the appropriate size 
and mass), and giving the bulge the appropriate projected effective 
radius from the spheroid size-mass relation in \citet{shen:size.mass} 
for its mass (knowing B/T), we can then construct a mock two-dimensional 
profile of the system. For simplicity, assume that the 
``bulge'' is spherically symmetric with a projected \citet{devaucouleurs} 
profile and the disk is azimuthally symmetric with 
an exponential surface brightness profile and vertical 
${\rm sech}^{2}$ profile (with the estimated $H/R$). 
Note that it makes no difference to our qualitative comparisons whether 
we adopt e.g.\ a pure exponential vertical profile instead, as 
suggested by \citet{barteldrees:exp.vertical.profiles,degrijs:vertical.disk.profiles}. 
Integrating through these 
mock profiles, we can construct the edge-on ``surface brightness'' 
map of the total system, and show this in Figure~\ref{fig:jz.distrib}. 
For both our estimate of disk heating here, and that in \citet{tothostriker:disk.heating}, we 
construct this toy model image for cases with net values 
$B/T=0.1-1$. 

The difference in galaxy shapes is immediately apparent. Again, the 
quantitative details will, in reality, reflect a much more complex history of the 
system. Here, we focus on two qualitative details. First, if heating were 
as efficient as the derivation of \citet{tothostriker:disk.heating} implies, there are essentially 
no very thin disks, even in systems with low $B/T$. This is similar to 
what we have found in \S~\ref{sec:consequences:mergerhistory} above -- the amount of 
heating for some merger history is, for realistic orbits, much less. 
Second, at intermediate $B/T$ typical of early-type spirals and S0 
galaxies, the derivation herein implies (owing to the low efficiency of 
disk heating) that disks, while clearly thicker, are still thinner than 
the bulge itself, giving galaxies a reasonably defined 
disk plus bulge or disk plus spheroid edge-on morphology, comparable 
to what is canonically observed along the Hubble sequence. 
On the other hand, if heating were linear in $\mu$, then it would 
more or less track $B/T$, giving thick disks 
(for $B/T\sim0.3-0.5$, this gives disk $H/R\sim0.3-0.5$, comparable to 
what is typically associated with galaxy bulges and extended spheroids, 
not ``disks'' in normal observational parlance). In fact, 
this is sufficient to largely suppress the ``disk plus bulge'' morphology 
entirely -- the Hubble sequence and changing bulge to disk 
ratios do not appear as such when viewed edge on, but rather (visually) 
would appear to be primarily a sequence in disk thickness. 

\breaker
\section{Discussion and Conclusions}
\label{sec:discussion}

We have shown that allowing for realistic satellite orbits -- in particular the fact 
that orbits are not exactly circular, and that satellites tend to be 
increasingly stripped as they execute multiple passages through a 
disk -- has dramatic implications for the efficiency and cosmological 
evolution of disk ``heating'' via minor mergers and galaxy-galaxy interactions. 

In previous generations of analyses
\citep[e.g.][]{tothostriker:disk.heating}, several assumptions were
made implicitly: that a satellite would enter a disk on a
slowly-decaying circular orbit owing to dynamical friction, and that
the satellite was rigid. In addition, much of this work focused on 
dissipationless (gas-free) cases. Under these conditions, it is
straightforward to derive the fractional disk
heating owing to a merger of a satellite of mass $M_{2}$ with a
primary of mass $M_{1}$, and to show that this heating (and
correspondingly the induced increase in disk scale height and vertical
velocity dispersion) scale linearly in the mass ratio, $\propto
M_{2}/M_{1}$.

This fundamental premise 
has largely informed the literature on the subject of disk heating 
and empirical constraints on merger histories from the thickness 
and vertical profiles of observed disks. However, we show 
that it is no longer the case when the simplifying assumptions are 
relaxed \citep[as in current, more detailed simulations and 
cosmological models; e.g.][]{velazquezwhite:disk.heating,
ardi:subhalo.mergers,benson:heating.model,
hayashichiba:subhalo.mergers,
kazantzidis:mw.merger.hist.sim}. 
These assumptions are fine when one considers a system 
that is effectively a point particle at large radius in an extended 
(but relatively low-density) background halo, on a moderately circular orbit, 
such that there are no close passages of the satellite and primary 
and the entire regime of the orbit has a characteristic frequency much 
larger than the internal characteristic orbital frequency of either the 
primary or secondary -- 
in the regime where the satellite is far from the galaxy and has just been 
accreted near the halo virial radius, for example. 
And in a strictly instantaneous sense, they are 
not bad. But in a global sense, in terms of estimating the effects on a galactic 
disk from a merger with initial mass ratio $M_{1}/M_{2}$, these 
assumptions break down, and the result is not just a numerical correction, but a 
qualitative change in the character of the heating. 

In modern $\Lambda$CDM cosmologies, 
orbits tend to be highly radial, especially those orbits 
that actually pass near or through the disk. Circular orbits efficiently lose 
angular momentum to interactions and other substructures in halo formation, 
and even when they survive, the dynamical friction times required for the 
orbit to decay from near the virial radius are so large (of order the Hubble time) 
that they tend to survive far away from the central galaxy rather than merge
with  
and/or heat the disk. Moreover, 
once the galaxy gets sufficiently close to the disk that it might effect some 
heating, the characteristic orbital frequencies are (by necessity) 
similar to the characteristic internal frequencies of the disk and halo. 
Because some of the galaxy will be in near-resonance with the 
secondary orbit, this allows much more efficient transfer of angular momentum 
from the orbital motion into internal motions, and even initially 
circular orbits become radial in just one or two passages. 
Because the orbital angular momentum is transferred 
quickly in a passage, the secondary 
cannot gently ``spiral in'' but must return from apocenter in a 
radial ``plunge'' -- the secondary effectively free-falls 
inwards (weakly decelerated by e.g.\ dynamical friction at this point) 
and ``punches through'' the primary, where it executes a rapidly damping 
oscillation through the primary on nearly radial orbits. At this stage, violent 
relaxation scatters stars off the time-dependent potential and transfers 
the orbital energy to the central stellar distribution (the bulge or whatever the 
hottest stars are at the center of the galaxy), damping the orbit and completing 
coalescence in a couple of dynamical times. At the same time, once 
the secondary begins to penetrate the primary disk, it experiences 
rapid tidal mass loss and soon most of the mass has been tidally added to the 
extended halo and cannot further heat the disk. 

For these reasons, the most useful approximation for disk heating 
in realistic cosmological situations is not a gradually decaying circular orbit 
of a rigid satellite, but rather something like the impact approximation: 
the heating is dominated by a single or a couple of radial oscillations at the initial 
mass, on a radial free-fall trajectory. We show that 
in this case, the heating is {\em not} linear in the mass ratio $M_{2}/M_{1}$. 
Rather, the effective heating is second-order in the mass ratio, 
$\propto (M_{2}/M_{1})^{2}$. 
We show that -- again for the reasons above -- this is a much more robust 
scaling. 

We demonstrate this by comparing our analytic predictions to a 
variety of numerical simulations, which include a range in 
secondary mass ratios, assumed structural properties of the 
disk, halo, bulge, and satellite, gas content, redshift, orbital parameters, 
and orbital anisotropy and angular momentum content. 
Remarkably, these different simulations, from various independent 
groups, all agree with our
analytic predicted scalings and with each other 
(when properly normalized) to within a factor $\sim2-3$. It is 
straightforward to show that a linear scaling in mass ratio is not a good description of the 
simulation results (even if the normalization of the linear scaling 
is free -- i.e.\ the scaling itself must be second-order). 

This provides a physical reason why simulations increasingly find, 
contrary to historical expectations, 
that systems can undergo cosmologically expected merger histories 
and still yield relatively thin disks at $z=0$ in line with 
observations \citep[see e.g.][]{velazquezwhite:disk.heating,
ardi:subhalo.mergers,abadi03:disk.structure,
benson:heating.model,
robertson:cosmological.disk.formation,robertson:disk.formation,
okamoto:feedback.vs.disk.morphology,hayashichiba:subhalo.mergers,
governato:disk.formation,kazantzidis:mw.merger.hist.sim,
younger:minor.mergers,villaloboshelmi:minor.mergers}. 
Many of these studies found 
that idealized numerical experiments yield less efficient disk 
heating than that predicted by \citet{tothostriker:disk.heating}; 
the key contribution of our work is to outline the physical reasons for 
this difference and to demonstrate that these differences are {\em not} 
a matter of e.g.\ the normalization or numerical prefactors 
involved in an estimate of disk heating: heating 
is second-order in mass ratio ($\mu$). 

This {\em qualitatively} 
changes the disk heating paradigm. 
Because the mass spectrum of mergers (in terms of halo mass) 
goes as ${\rm d}N/{\rm d}\ln{\mu}\propto \mu^{-1}$, 
if the heating per encounter were in fact linear ($\propto \mu$), 
then there would be more or less equal contributions to 
disk heating and thick disk formation from each logarithmic interval 
in mass ratio. 
Observationally, this would imply that no more than a few dwarf galaxies 
with properties similar to those of e.g.\ the rather small Sagittarius dwarf 
\citep[see][]{helmi04:sgr.stream,fellhauer06:sgr.stream.orbit} 
could have been accreted or destroyed by the Milky Way in the last 
10\,Gyr, in conflict with cosmological expectations 
and making the well-known ``missing satellite'' problem even 
more difficult to resolve. 

In contrast, if the heating is second-order in $\mu$, 
then the net heating is dominated by the few large mergers 
of relatively large mass ratios. 
This behavior has in fact been recognized in cosmological simulations, 
and yields better agreement with observations \citep[see][]{
abadi03:disk.structure,
robertson:cosmological.disk.formation,benson:heating.model,
okamoto:feedback.vs.disk.morphology,
governato:disk.formation,kazantzidis:mw.merger.hist.sim,read:thick.disk.cosmo.sims}. 
It also fundamentally enables the existence of very thin disks: 
since mergers of relatively large mass ratios $\mu\gtrsim0.1$ 
are rare, they are somewhat stochastic -- i.e.\ some systems 
which have gone sufficient time or been sufficiently gas rich 
since their last $\sim$\,1:10 merger will exist and can therefore have 
very thin disks ($H/R\lesssim0.1$) 
similar to the ``ultra-thin'' or ``super-thin'' disks observed 
\citep{seth:disk.scale.heights,
yoachim:disk.scale.heights}. 
If instead the same net heating came from mergers 
of mass ratios $\mu\sim0.001-0.01$ (where the expected number of 
mergers since $z\sim1$ is of order several hundred per galaxy; 
making the existence of any galaxies without such mergers in their 
recent past an improbable event) as from mergers 
with mass ratio $\mu\sim0.1-1$, then it would be almost impossible to 
contrive realistic cosmological merger histories that could give rise to 
ultra-thin disks.

There are a number of other indirect implications of this result. 
For example, we demonstrate that this is of significant importance for 
the edge-on morphologies of disk plus bulge systems. Given 
some bulge formation in mergers, 
whether the induced heating in the surviving disk goes linearly or 
second-order in $\mu$ produces a different relation between 
scale heights and bulge-to-disk ratios and, correspondingly, 
different morphologies in observed early-type disks. 
If realistic scale heights increased as efficiently as expected in the 
circular orbit, rigid secondary approximation, then disks of intermediate 
$B/T$ would necessarily be thicker than observed -- in fact, the Hubble sequence, viewed 
edge-on, would no longer have a characteristic ``bulge plus disk'' morphology 
and would instead appear as a sequence in ellipticity of characteristically boxy 
isophotal shapes (because a sufficiently thick disk 
appears in edge-on projection as somewhat rectangular). 
In contrast, disks appear to remain thin even in systems with significant 
(or dominant) bulges, giving rise to characteristically ``disky'' 
(``lemon-shaped'') isophotal shapes. The non-linearity of disk heating 
can explain why galactic morphological components 
form bimodal families (bulges and disks), as opposed to 
a sequence of increasingly ``spheroidal'' disks. 

Applying these recalibrated models to the Milky Way and other massive disk 
galaxies, the implied constraints on the merger history from the thick 
and thin disk are dramatically changed and fully consistent with cosmologically 
expected merger histories. In detail, applying our scaling, calibrated against 
numerical simulations, to the observed age-velocity dispersion 
relation in the solar neighborhood, implies that the Milky Way could have 
survived as many as $\sim5-10$ minor (mass ratio $\sim$\,1:10) mergers 
in the last 10\,Gyr. If some of the disk thickness owes to other processes or 
there were some more violent events, this number might be a bit smaller, 
but if the Milky Way were more gas-rich in the past, then this 
number would in fact go up by a factor $1/(1-\fgas)$. In either case, the median 
number of such mergers expected for a Milky-Way mass halo is only 
$\sim3$ (with $\sim90\%$ of Milky-Way like systems having $<5-6$ such 
mergers in this interval). The observed vertical velocity dispersions of stars (and implied 
disk thickness) of the Milky Way is completely consistent with the cosmologically 
expected amount of minor merging in the last 10\,Gyr, and in fact these 
constraints alone permit the 
Milky Way, far from being especially quiescent, to be in the upper 
cosmological quartile in 
terms of number of minor mergers since $z\sim1-2$. 
Translating these constraints to a ``maximum'' mass ratio merger 
implies that the Milky Way could even have survived mergers of mass 
ratio $\sim$\,1:4-1:3 since $z\sim2$, without producing too much heating or 
thick disk. 

Because the heating is second order in mass ratio, 
there is almost no measurable heating effected by encounters
with mass ratios 
$\mu \ll 0.1$, so there is no overall constraint 
based on the thickness of the disk or internal velocity dispersions 
on how much {\em total} mass has been added to the Milky Way via merging. 
In contrast to the $<5\%$ growth since $z\sim2$ inferred by 
\citet{tothostriker:disk.heating}, we find that the Milky Way could have accumulated 
up to $\sim25-50\%$ of its present mass in the last 10\,Gyr by 1:10 mergers alone 
($\sim5-10$ individual 1:10 mergers). More interesting, because again smaller 
mass ratio mergers heat only to second order in $\mu$, {\em the Milky Way could 
easily, in fact, have grown by factors of several since $z\sim2$}, provided 
this growth mostly came from relatively small mass ratio mergers 
$\mu \lesssim$\,1:10.
This is critical for our understanding of galaxy formation: 
at some level of sufficiently small mass ratios below $\sim$\,1:10-1:100, 
the distinction between accretion and mergers is increasingly blurred. 
Essentially all galaxies are expected to 
gain a large fraction of their mass and grow by large factors 
(up to an order of magnitude) in the last $\sim10\,$Gyr from such accretion 
events. If these small events heated as efficiently as major events 
(e.g.\ if $\sim30$ separate 1:100 mergers were as damaging to a disk 
as a 1:3 merger), there would be a fundamental difficulty for 
CDM models. In fact, owing to the inefficiency of disk heating, 
there is no tension here. 

This is not, of course, to say that problems of disk formation 
and survival in a $\Lambda$CDM Universe have been solved. 
There are still well-known problems related to the loss of angular 
momentum in disks (owing in part to over-merging), 
that tend to yield (in cosmological simulations) disks that 
are too concentrated and have too much bulge 
relative to observations. A great deal of theoretical 
effort has gone into studying this \citep{weil98:cooling.suppression.key.to.disks,
sommerlarsen99:disk.sne.fb,sommerlarsen03:disk.sne.fb,
thackercouchman00,thackercouchman01,abadi03:disk.structure,
governato04:resolution.fx,governato:disk.formation,
robertson:cosmological.disk.formation,
okamoto:feedback.vs.disk.morphology,scannapieco:fb.disk.sims}, 
with increasing agreement that some combination of 
improved resolution and proper implementation of feedback from 
supernovae and stellar winds are critical for the formation of 
realistic disks. These issues are not addressed or affected by 
our analysis here.

However, it is increasingly clear 
(as resolution improves such that thick disks can be properly 
resolved in the vertical direction) 
that these simulations do {\em not} generally have the 
problem of producing disks that are much 
too thick relative to observations. In this paper, we have shown that there is a 
good physical reason for this, and that it is not a peculiarity of a 
couple of simulations or of an especially quiescent merger 
history, but rather is a robust and quite general expectation even 
for systems with violent merger histories and even in systems 
that will be bulge-dominated at $z=0$. We have derived a new, 
more generally applicable scaling of disk height with 
merger mass ratio, and shown that it 
qualitatively changes the expectations for how disk 
heating should proceed in a $\Lambda$CDM universe, and 
generically leads to the expectation that disks are quite robust 
to being heated by minor mergers. Although there are still 
considerable improvements in models needed to explain 
when and how disks form and why the distribution of bulge-to-disk 
ratios is what is observed at $z=0$, there is 
no tension between the models and the heating and thickness of disks. 
In short, we have argued 
that there is no ``disk heating'' problem in $\Lambda$CDM 
cosmologies, once a proper accounting of disk heating 
for realistic cosmological orbits and conditions is considered.

\acknowledgments 
We thank Shardha Jogee, John Kormendy, and 
Todd Thompson for helpful discussions. 
We also thank Kyle Stewart for sharing suggestions and simulation 
results. This work
was supported in part by NSF grants ACI 96-19019, AST 00-71019, AST
02-06299, and AST 03-07690, and NASA ATP grants NAG5-12140,
NAG5-13292, and NAG5-13381. Support for 
TJC was provided by the W.~M.\ Keck 
Foundation. 

\bibliography{/Users/phopkins/Documents/lars_galaxies/papers/ms}

\end{document}